\documentclass[english,aps,prstper,reprint,showpacs,titlepage,longbibliography]{revtex4-2}   

\usepackage[T1]{fontenc}
\usepackage[latin9]{inputenc}
\usepackage{geometry}
\usepackage{graphicx}
\usepackage[above,below]{placeins}
\usepackage{times}
\usepackage{hyperref}  
\hypersetup{colorlinks=true,urlcolor=blue,citecolor=blue,linkcolor=blue}   
\usepackage{enumitem}         
\setlist{nosep}                 

\usepackage{graphicx}
\usepackage{subcaption}
\usepackage{dcolumn}
\usepackage{tabularx}
\usepackage{multirow}
\usepackage{bm}
\usepackage{amsfonts,color}
\usepackage[table,xcdraw]{xcolor}
\usepackage[english]{babel}
\usepackage{lipsum} 
\usepackage{array}

\definecolor{Gray}{gray}{0.9}

\newcolumntype{Y}{>{\centering\arraybackslash}X}
\newcolumntype{Z}{>{\raggedright\arraybackslash}X}

\newcolumntype{P}[1]{>{\raggedright\arraybackslash}p{#1}}

\begin{document}

\begin{titlepage}

\title{Student Perspectives on Traditional Pedagogy Used in Graduate Physics Coursework}

\author{Kevin Coldren (he/him)}
 \affiliation{School of Physics and Astronomy, Rochester Institute of Technology, Rochester, NY 14623}

\author{Nkodia Ngondala (he/him)}
\affiliation{School of Physics and Astronomy, Rochester Institute of Technology, Rochester, NY 14623}

\author{Audrey Claar (she/her)}
\affiliation{School of Physics and Astronomy, Rochester Institute of Technology, Rochester, NY 14623}

\author{Mike Verostek (he/him)}
\affiliation{School of Physics and Astronomy, Rochester Institute of Technology, Rochester, NY 14623}

\author{Diana Sachmpazidi (she/her)}
\affiliation{School of Physics and Astronomy, Rochester Institute of Technology, Rochester, NY 14623}

\begin{abstract}
Graduate programs in STEM disciplines are central to preparing future researchers and professors. Program requirements for students often include taking several graduate-level courses. Anecdotal evidence suggests that graduate coursework in physics in particular features outdated and ineffective pedagogical methods, with high emphasis on mathematical rigor in place of conceptual learning or connections made with authentic research. Prior physics education research indicates that students can leave graduate courses with shortcomings in conceptual understanding of covered topics. This pilot study is designed to document student perspectives on their graduate coursework in a single U.S. R1 physics Ph.D. program. A total of 14 semi-structured interviews were conducted with students enrolled in the program, and thematic analysis was conducted on five of these interviews for this paper. The resulting themes are discussed, including the prevalence of traditional passive lecture pedagogy and students placing high value on course content relevant to research.

\clearpage
\end{abstract}

\maketitle
\end{titlepage}

\section{\label{sec:Introduction }Introduction}

Graduate programs in STEM fields play an essential role in preparing the next generation of researchers, educators, and scientists \citep{national2018graduate}. In physics, graduate school is often viewed as the main path for students pursuing research, academia, or other advanced scientific careers. Because of this, graduate training matters. It shapes not only what students know, but also how they see themselves as physicists and how prepared they feel to continue in the field.

At the same time, graduate education faces ongoing challenges. STEM graduate programs continue to struggle with retention, diversity, and student wellness \citep{posselt2021promoting, sowell2015doctoral}. Many graduate students experience stress, anxiety, depression, isolation, and uncertainty about whether they belong in their programs \citep{posselt2021promoting, stachl2020sense}. These issues are especially important in physics, where the field's culture has historically been shaped by narrow ideas of who belongs in science and what it means to be successful \citep{stachl2020sense, sachmpazidi2025role}. For many students, graduate school is both an academic and social challenge.

Much recent research on graduate education has focused on departmental culture and structure, including advising, mentoring, belonging, climate, expectations, and student support \citep{posselt2021promoting, stachl2020sense, sachmpazidi2025role}. This work shows that graduate education is not only about individual effort or intelligence. Students' experiences are also shaped by the systems, relationships, and cultures around them.

However, less attention has been given to the classroom and coursework experiences of graduate students, especially during the first two years of graduate school. These early years usually include required courses, teaching assistant responsibilities, qualifying exams, and the transition into research. This period can strongly influence how students understand physics, how confident they feel, and whether they see themselves developing as researchers \citep{sachmpazidi2025role}. It is also important because much doctoral attrition happens during the early years of graduate programs \citep{sowell2015doctoral}. Even so, graduate coursework is often treated as something students must simply get through, rather than as a learning environment that should be studied and improved.

Physics education research has studied teaching and learning at the undergraduate level for decades \citep{docktor2014synthesis}. Prior research has shown that traditional lecture-based instruction does not always lead to deep conceptual understanding, even when students can complete advanced calculations \citep{brundage2023using}. In response, researchers have developed active learning strategies, research-based instructional materials, and other approaches that help students think deeply, collaborate, and connect concepts for problem-solving \citep{meltzer2012resource}. These methods have been shown to support learning in undergraduate and high school classrooms \citep{meltzer2012resource, sundstrom2025relative}. This raises the question of whether similar approaches could also support learning at the graduate level.

Despite this, anecdotal evidence suggests that graduate physics course are typically taught in a traditional lecture approach that emphasizes mathematical strictness, procedures, and individual performance. However, there is currently little empirical evidence to document graduate course pedagogy and student experiences of the coursework. While rigor remains important in physics, rigor alone does not always lead to meaningful learning. Graduate students may still struggle to connect coursework to conceptual understanding, research practices, or their own development as physicists \citep{robbins2025decision}. These courses can also reinforce the idea that physics is something students are expected to figure out alone, rather than through collaboration, discussion, and support.

Some research has examined graduate-level physics pedagogy, including graduate students' conceptual understanding, problem-solving, and learning in graduate physics courses \citep{robbins2025decision, singh2008student, mason2010surveying}. These findings suggest that graduate students, like undergraduate students, benefit from instruction that makes thinking visible, addresses conceptual difficulties, and connects coursework to actual physics work \citep{robbins2025decision, brundage2023using, lepage2021active, porter2020effectiveness}. Still, more research is needed to understand how graduate students experience their coursework and how those experiences affect their learning, confidence, and sense of belonging.

This pilot study examines the coursework experiences of graduate students in a single R1 physics Ph.D. program in the United States. The goal is to understand how students describe the teaching, structure, and learning environment of their required graduate courses. To do this,  semi-structured interviews were conducted with graduate students currently enrolled in the program. The interview transcripts were analyzed using thematic analysis \cite{BraunClarke2006}, and the themes that emerged are discussed in this paper. By focusing on graduate students' own perspectives, this study contributes to conversations about graduate education, physics pedagogy, and how doctoral programs can better support students during the early years of graduate training. The work done for this paper was guided by the following research questions:

\textit{RQ1: What are student perspectives on the pedagogical methods used in their graduate physics courses?}

\textit{RQ2: What course learning outcomes do students value in their graduate physics courses?}

\section{\label{sec:Methods}Methods}
\subsection{Educational context}
This study was conducted in a physics Ph.D. program at an R1 university in the United States with a typical population of approximately 100 students. Coursework requirements for this program include four each of required and elective graduate courses. These courses are typically taken in the first two years of the program, with the first year including all four required courses and two electives. Students then transition to full-time research toward their eventual dissertation in their second year while finishing their remaining electives. Required courses include standard graduate physics courses (e.g. quantum mechanics, electrodynamics, etc.) while the elective courses have more variety in subfields of physics.

\subsection{Procedures and Analysis}

Participants were graduate students who were enrolled in the Ph.D. program at the time of data collection in June of 2025, and were recruited via email. 30-minute semi-structured interviews were conducted remotely via Zoom with a total of 14 students. Participants were at a variety of stages in the program, ranging from the end of the first year to seventh year and beyond. Interview transcripts were then cleaned and checked for accuracy.

For the purposes of this paper, a subset of five student interviews were chosen for thematic analysis. These  students are referred to by the following pseudonyms: Andrew (Year 1), James (Year 1), Riley (Year 2), Michael (Year 4), and Ashley (Year 5). This subset was chosen to have perspectives from students at a variety of stages of in the program. 

Thematic analysis was conducted on the interview transcripts with a combined deductive-inductive approach. A codebook was developed from the initial interview protocol, then applied to one interview transcript in a first-pass coding process. After discussions between the co-authors, a second version of the codebook was then developed. This version of the codebook was then applied to all five transcripts in the qualitative software Dedoose, with memos attached to each coded excerpt. The excerpts, memos, and codes were then imported into the virtual white boarding software Lucid~\cite{Lucidchart2026}. The data were then organized into emerging themes, which are discussed below. 

\subsection{Limitations}
Inter-rater reliability  has not yet been conducted on the codebook used in this study, but is planned for the future. The five student interviews used for this paper are a small sample size for both our total collected data as well as for the student population in this program as a whole. This paper also only captures the student perspective in this context and not those of other stakeholders like program leadership and teaching faculty. The nature of this pilot study limits the claims that can be made about the broader landscape of graduate physics coursework, but does show the need for larger future studies.

\section{\label{sec:Results}Results}
Two major themes emerged from the experiences of the five students in their graduate coursework. The first theme mainly dealt with the perceived effectiveness of pedagogical methods used by professors. The second theme centered around student views on the value of their coursework toward their development as researchers and scientists. 

\subsection{Most Courses in This Program Feature Passive Lecture Pedagogy with Wide Variation in Student Learning Outcomes}
When asked how a typical graduate physics course in this Ph.D. program was run, all five of the interviewed students responded that both required and elective classes were taught in a traditional lecture model with minimal deviations. Instructors would write information on the board with accompanying verbal explanations, while students passively receive this information. The only significant deviation was in two astrophysics electives taken by Andrew which featured some discussion-based classes centered around research topics. 

Most of these classes placed a heavy emphasis on mathematical skills, usually assessed through homework assignments and summative written exams, though some courses did have final projects as an alternative. The emphasis on mathematics is shown in the experience Michael had in a plasma physics elective course:

\begin{quote}
``[The class] was just boring. [...] I don't need to see these long equations where [it] stretches all the way across the 4 segment board. [...] I could have given that a pass if we were given any motivation for why you should care about the result of the derivation.'' - Michael, Year 4
\end{quote}

Here Michael is expressing frustration with spending such a large portion of a lecture on mathematical derivations with little connections made to physical concepts or understanding, leading to a lack of engagement. 

When discussing positive experiences in their classes, the most common response was centered around a few professors who would post well-organized notes before the lecture on their subject, as Andrew describes:

\begin{quote}
``I had two professors that would always post [class notes] before the lecture, so that you could be prepared. [...] I'm the kind of person that wants to read the notes before I go [...] I'm not gonna read them afterwards, because I was already there.'' - Andrew, Year 1
\end{quote}

Here Andrew is exemplifying the common sentiment that the notes posted by these professors functioned as a way to ensure that the lecture was not the first time the students were processing the material, without the need to read the textbook for the course. At the same time, some interviewed students used this as a way to be even more passive during class. Students also felt that the courses that featured this structure had overall better organization and clarity than others, as discussed by Michael:

\begin{quote}
``Stat mech and condensed matter felt similar to me, where I did know what I was supposed to be learning, and why I was supposed to be learning it. That's [better] than `We're going to
[do] this derivation, because once it's done, we'll [...] feel proud of ourselves.' '' - Michael, Year 4
\end{quote}

Here Michael is juxtaposing his more positive experiences in well-organized courses with those in classes that were not. These two courses are typically taught by a specific professor that was consistently praised by all five students, mentioning the overall organization, useful course notes, and instructor questions during lectures to engage students who read them beforehand. This professor has taught the required electrodynamics course in addition to statistical mechanics for several years, which may be a contributing factor to the overall course structure, instructor familiarity with the material, and common student struggles with the subject matter.

A common topic of discussion around course lectures was professor responsiveness to students during class lectures. Some students had generally positive feelings on this topic:

\begin{quote}
 ``I have not encountered a class where the professor genuinely just stands there and talks at you for 90 minutes. There is always at the minimum [some] checking to make sure that people are paying attention.'' - Andrew, Year 1
\end{quote}

Riley had a more mixed sentiment:

\begin{quote}
``I feel like often the best professors are the ones that notice when everyone is not understanding something. Because that is certainly [something] I've seen go both ways.'' - Riley, Year 2
\end{quote}

Ashley had some more negative experiences in this vein:

\begin{quote}
``You can have professors who will look at you occasionally, but still spend minutes at a time with their back to you, and you have a question that never gets answered.'' - Ashley, Year 5 
\end{quote}

These three quotes show the variety of instructional quality that these students have experienced in their collective time spent in graduate coursework. Some professors are actively looking for student confusion and attempting to address it, while others are not putting as much effort into informal formative assessment during their lectures.

The interviews analyzed for this paper often focused on negative experiences had by the students in their graduate physics coursework. All five students described at least one course where they felt minimal learning occurred, and both elective and required coursework were mentioned. For example, James described his time in a plasma physics course:

\begin{quote}
``I would never go near [another] one of his classes for any amount of money, because that was intensely frustrating.[...] The impression I had was that he was so familiar with the material he'd forgotten what was not obvious to somebody who wasn't studying plasmas for 40 years.'' - James, Year 1
\end{quote}

James is highlighting the fact that this professor was not able to relate high-level physical concepts to the level of a graduate level student. A similar story comes from Michael's experience in a quantum dynamics course:

\begin{quote}
``Quantum dynamics was the worst time I've ever had in my life. [...] The professor was completely disconnected from the understanding that the students had. [...] There were a couple lectures where he had read the material a few times, and there 
[were parts] that he fundamentally didn't understand.'' - Michael, Year 4
\end{quote}

This example from Michael shows that some professors in this program are teaching graduate courses that they may lack the preparation to teach, whether in terms of content knowledge, available time, or prioritizing other areas. Ashley agreed with this sentiment: 
    
\begin{quote}
``In my opinion, they seem like they're at this university to do research, and they are forced to teach. So they're teaching, but they're not gonna do anything extra.'' - Ashley, Year 5
\end{quote}

Ashley seems to have a perception that professors at this university prioritize other aspects of their jobs over their teaching assignments. This is part of a trend in the student data that some of their course lectures are of poor quality, featuring unclear or rushed explanations of the course topics. 

\subsection{Students Placed the Highest Value on Course Outcomes Related to Research Preparation}
The interviewed students consistently placed high value on courses and learning outcomes that were directly relevant to the research they were involved with toward their eventual dissertations, as summarized by Riley here:

\begin{quote}
``Yeah, beyond the intro classes the most useful classes [...] are ones where I can see some applicability to my own work.'' - Riley, Year 2
\end{quote}

Riley here was mostly referring to elective courses specific to their chosen subfield of physics, as did most of the students when talking about course relevance to their research. However, James discussed this concept when describing the engaging aspects of his required quantum mechanics course: 

\begin{quote}
``I think the thing that I liked about his class is that every one of his problems [...] related to something that was either being studied or in a paper. [...] I still had to do the same [math], but it was really interesting.'' - James, Year 1 
\end{quote}

While the research that James does was not directly related in this case, he still found the authentic real-world examples used by his professor to be engaging and meaningful.  

Students also identified other outcomes from their coursework as having value, which were mainly described as learning general skills relevant physics in general. When asked if his courses helped prepare him for his research, Andrew said:

\begin{quote}
``I wouldn't say that they necessarily align with it, but I would say that they just cover basic concepts that are kind of relevant to everything. [...] I also learned how to code.'' - Andrew, Year 1
\end{quote}

Andrew here is expressing his view that there are valuable learning outcomes from his courses that were not necessarily directly related to his research, but rather more general skills and knowledge that may be useful down the road. 

When discussing his transition to his dissertation research after completing his coursework, Michael reflected on what might not have worked for him:

\begin{quote}
``I've learned as I'm doing research [that] there are a couple areas where I'm missing some foundational information, and [...] filling that in on your own without guidance is way harder.'' - Michael, Year 4
\end{quote}

Michael's sentiment here indicates that at some level, his coursework did not fully prepare him for his research, which he saw as a shortcoming. This shows that student ability to effectively and efficiently conduct research toward their dissertation is generally their highest priority. 

Students also had nuanced views toward the amount of time spent on coursework in their program. When discussing the utility of what she learned in her courses, Ashley said:

\begin{quote}
``Honestly, [on] a day to day, I don't really use any of it. [...] I have gone back to the particle physics [electives] as I'm writing the intro to my thesis, and I have to re-remember stuff, so I think those were useful courses.'' - Ashley, Year 5
\end{quote}

Ashley here is only placing value on coursework that is directly useful and applicable for her dissertation, while also viewing any other courses as wasted time toward her graduation. This is part of a larger trend in the student data on the conflict between the amount time spent on coursework against time spent toward eventual graduation. Ashley came into her program with a research area in mind, and thus she felt that in an ideal scenario she would have just taken particle physics courses instead of the more general required courses. Michael also shared this sentiment toward his own graduate courses. However, both also acknowledged this as unrealistic, as exemplified here by Ashley:

\begin{quote}
``So maybe we need to change what coursework looks like in grad school. But if we do that, then you really have to come in knowing what you want to do, and I don't think we should force anyone to do that either.'' - Ashley, Year 5
\end{quote}

Ashley is acknowledging that not all students come into graduate school with the same background and motivations that she did. Using coursework as a way to explore physics subfield for potential research pathways was mentioned by other students in our data:

\begin{quote}
``Exposing people to different kinds of physics, the easiest way of exploring [subfields] is taking a bunch of courses and seeing what things exist and what they're applied to. '' - Riley, Year 2
\end{quote}

Riley's sentiment was true for Andrew and James as well. Andrew was able to explore subfields in his first year and find something that interested him. On the other hand, James was pushed toward theoretical physics due to a primarily negative experience in his required data science course. 

\section{\label{sec:Discussion & Conclusions} Discussion \& Conclusions}
The analysis presented in this paper shows that the students in this program take courses that are largely taught using passive lectures that emphasize mathematically rigorous derivations while offering few student-centered teaching techniques. In addition, students mostly held the viewpoint that only course learning outcomes directly related to their dissertation research were worth the considerable time and effort required both in and out of class. Documentation of these classroom experiences is largely absent from the existing literature on graduate physics education.

Well-documented research in the field of undergraduate STEM education on the effectiveness of active learning teaching methods over passive lectures~\cite{freeman2014active} suggests that the teaching methods employed in the majority of graduate courses in this program are outdated and ineffective toward student learning. If active learning techniques were to be considered in this context, significant course restructures like studio physics or flipped classroom instruction can have logistical or financial barriers to their implementation~\cite{borte2023barriers}. However, recent research suggests that lecture-based courses that incorporate a moderate amount of active learning techniques like clicker questions, discussion, and group problem-solving lead to significant learning gains for students~\cite{ross2026predictive}. Alignment between departmental leadership and teaching faculty would be critical in the  implementation of these teaching methods~\cite{white2016adopting}.

The emphasis placed on course outcomes related to research skills by students is consistent with the amount of time and effort spent on work toward a dissertation in a physics Ph.D. program. Other outcomes like physics content knowledge were not as highly valued by these students, especially if not actively used in their daily work as researchers. This may be
evidence of a disconnect in expectations for graduate coursework between students and program faculty, and is something for physics Ph.D. programs to take into account when designing coursework for their students.

This paper represents an in-progress research project, as these methods will be applied to the remainder of the 14 collected student interviews from this program, and the completed analysis will be presented in a future publication. Documenting student experiences at other institutions, along with perspectives from program faculty and leadership, will be major features of future studies in order to paint a more complete picture of the landscape in graduate physics pedagogy.

%

\end{document}